# Large gap, a pseudogap and proximity effect in the Bi$_2$Te$_3$/Fe$_{1+y}$Te interfacial superconductor

M. Q. He[1], Q. L. He[1], J. Y. Shen[1], H. C. Liu[1], Y. Zheng[1], C. H. Wong[1], Q. H. Chen[1], J. N. Wang[1], K. T. Law[1], I. K. Sou[1], A. P. Petrovic and R. Lortz[1,*]

[1]Department of Physics, The Hong Kong University of Science and Technology, Clear Water Bay, Kowloon, Hong Kong S. A. R., China.

[*]Corresponding author: lortz@ust.hk

We report directional point-contact spectroscopy data on the novel Bi$_2$Te$_3$/Fe$_{1+y}$Te interfacial superconductor for a Bi$_2$Te$_3$ thickness of 9 quintuple layers, bonded by van der Waals epitaxy to a Fe$_{1+y}$Te film at an atomically sharp interface. Our data show a very large superconducting twin-gap structure with an energy scale exceeding that of bulk FeSe or FeSe$_{1-x}$Te$_x$ by a factor of 4. While the larger gap is isotropic and attributed to a thin FeTe layer in proximity of the interface, the smaller gap has a pronounced anisotropy and is associated with proximity-induced superconductivity in the topological insulator Bi$_2$Te$_3$. Zero resistance is lost above 8 K, but superconducting fluctuations are visible up to at least 12 K and the large gap is replaced by a pseudogap that persists up to 40 K. The spectra show a pronounced zero-bias conductance peak in the superconducting state, which may be a signature of an unconventional pairing mechanism.

Tailoring a topological superconducting material by combining the topologically protected surface states of a TI with a superconductor via the proximity effect is of enormous theoretical and technological interest, principally due to the possibility of finding the Majorana fermion states, which are expected to exist in the vortex cores of topological superconductors [27, 28]. We have recently reported 2D superconductivity at the interface of Bi$_2$Te$_3$/Fe$_{1+y}$Te heterostructures below 12 K [25]. Although both parent materials are non-superconducting, the interface becomes a 2D superconductor. Bi$_2$Te$_3$ is a 3D topological insulator (TI) whose surface states consist of a single Dirac cone at the Γ point [26], while Fe$_{1+y}$Te is the parent compound of a '11' family of Fe-based superconductors. Despite the exact mechanism for superconductivity in Bi$_2$Te$_3$/Fe$_{1+y}$Te remaining unknown, it has been shown that any doping effect by O, Bi or Te impurities can be excluded [25]. Evidence that TI surface states play a role in the emergence of superconductivity is found in the fact that a critical thickness of the Bi$_2$Te$_3$ layer is required [25]. As the number of Bi$_2$Te$_3$ quintuple layers (QL) increases, $T_c$ rises from ~1.2 K (1 QL) to 12 K (5 QL) and then saturates. This correlates with the 5 QL critical thickness required to form a fully-developed surface state in Bi$_2$Te$_3$ [29, 30]. It is therefore of particular relevance to determine whether proximity-induced topological superconductivity exists in the Bi$_2$Te$_3$ layer of this heterostructure.

In this letter, we report directional point-contact data measured for current injection (a) parallel to the interface into the edge of the heterostructure, simultaneously probing both layers and (b) perpendicular to the interface into the top Bi$_2$Te$_3$ layer. Point-contact spectroscopy is an energy-resolved technique directly probing the amplitude, symmetry and temperature-dependence of the superconducting gap [31]. Our data were acquired using bilayers of Bi$_2$Te$_3$(9QLs)/Fe$_{1+y}$Te, chosen for their high $T_c$ = 12 K. The Van-der-Waals bonding between the materials results in extremely high quality atomically-sharp interfaces; more growth and structural characterization details may be found in Ref. [25]. To fabricate a point-contact device on the edge of the bilayer, a thin slab was glued onto a silicon substrate with one edge facing upwards. Ordinary contacts with resistance of several Ohms were prepared by silver paint on the $Bi_2Te_3$ surface. The edge of the sample was finely polished before a thin layer of Au was deposited onto the edge. Subsequently, an isolated 100 nm wide Au strip was separated using a focused-ion-beam. The maximum contact area is ~100 × 149 nm$^2$ (from the width of the Au strip and the total thickness of the FeTe/$Bi_2Te_3$ bilayer, respectively). To also achieve perpendicular current injection with the aim of highlighting any proximity-induced topological superconductivity in the $Bi_2Te_3$ layer, we prepared another point contact device by gently pushing a tungsten scanning probe tip onto the $Bi_2Te_3$ layer. All data reported in this work were acquired on contacts of a few kΩ and the data reproducibility was verifies on 4 different point-contact devices. The point-contact spectral shape is strongly dependent on the tunnel barrier height parameter $Z$ [31], which is linked to the contact resistance. $Z = 0$ corresponds to pure Andreev reflection, while larger values represent the spectroscopic tunneling regime. Achieving $Z > 0$ is crucial for determining the low-energy gap structure; we consistently achieve $Z \geq 0.35$ in our devices. Furthermore, our nanoscale contact area ensures that our experimental tunneling regime is ballistic and not thermal or diffusive, i.e. the applied bias voltage $V_b$ corresponds to the electron injection energy. This is confirmed by the temperature-independent value of the normal-state contact resistance.

The differential conductance $dI/dV$ vs $V_b$ was measured at temperatures from 0.27 K to 70 K in magnetic fields up to 15 T with a quasi-four-probe method, using a Keithley 6221 AC/DC current source to generate a small, constant-amplitude (10 nA) AC current $I_{AC}$ with frequency 5 Hz, superposed on a ramped DC bias current. A standard lockin-technique in combination with a DC multimeter was used to measure $dI/dV$ and $V_b = V_{DC}$ across the junction.

Fig. 1a shows the temperature dependence of the point-contact spectra for the nanocontact on the edge of the heterostructure. At the highest temperatures a smooth parabolic background is seen, with little difference in data acquired at 70, 50 and 40 K. Below 40 K a pseudogap develops symmetrically around $V_b = 0$, gradually deepening as the temperature falls. Until 15 K, the gap is rounded at low energy, but at 12K the conductance flattens around $V_b = 0$ prior to the emergence of a ZBCP below ~10 K. Concurrently, shoulder-like structures develop at ~10 mV and ~5 mV, which as we will now demonstrate correspond to a phase-coherent superconducting twin-gap structure [34].

To fit the temperature dependence of our data, we primarily employ a modified Blonder-Tinkham-Klapwijk (BTK) model for finite-transparency tunnel junctions [35]. Our fits are based on a 1D BTK model for simplicity, since higher-dimensional models are equivalent to the 1D case except for small shifts in the barrier heights $Z$. In such a highly two-dimensional superconductor, fluctuations are expected to significantly reduce the quasiparticle lifetime; we account for this in our model with an energy-dependent Dynes parameter [36] of the form $\Gamma \exp[(|V_b|-\Delta)/W]$ where $\Gamma$ and $W$ are free parameters. The shoulders at ~5 and ~10mV are modeled by an anisotropic two-band s-wave order parameter $\Delta(\alpha+(1-\alpha)\cos\theta)$ [37], in which we determine the gap anisotropies $\alpha_{1,2}$ from a fit at our lowest achievable temperature 0.27K, then fix $\alpha_{1,2}$ at these values for all other temperatures.

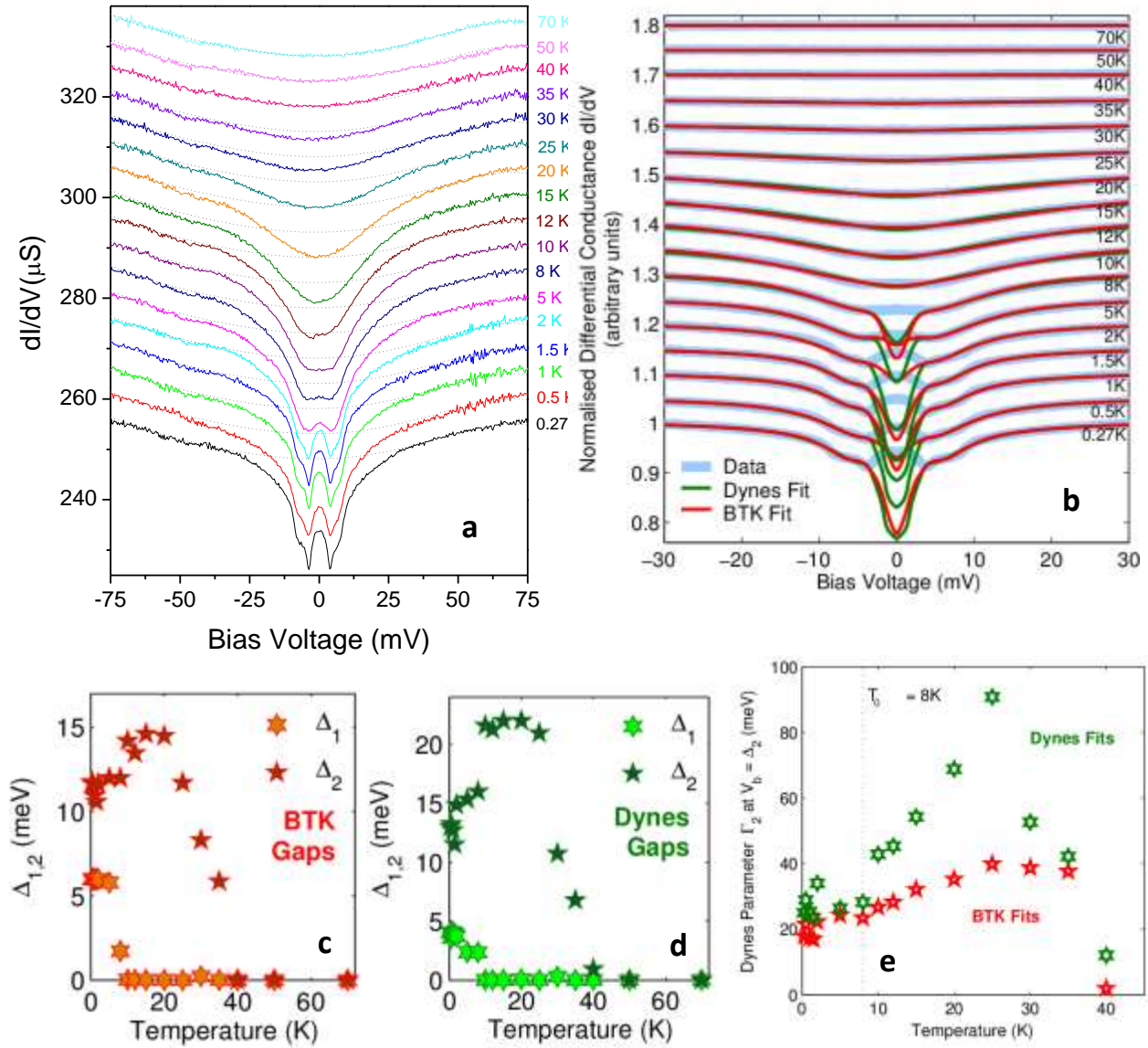

FIG. 1. **a** Temperature-dependent differential conductance spectroscopy of a point-contact on the edge of the $Bi_2Te_3$(9QLs)/$Fe_{1+y}$Te heterostructure. An offset has been added to all data except the 0.27 K spectrum for clarity. The dotted lines illustrate the normal-state background as obtained from a polynomial fit to the 70 K data. **b** Data from 2a normalized to a polynomial background from the 70K spectrum, together with fits using superconducting 2-gap BTK and Dynes models. Each spectrum is offset by 0.05 from the previous one for clarity. The low voltage range containing the ZBCP (±3.5 mV) was excluded from the fit. **c,d** Temperature dependence of the two superconducting gaps $\Delta_{1,2}$ (which close at ~8 K and 40 K respectively) determined from BTK and Dynes models. **e** The energy dependent Dynes parameter $\Gamma_2$ obtained from the fits to the 2-gap Dynes and BTK models describing the quasiparticle lifetime in $\Delta_2$.

The fitting results are shown in Fig. 1b: we are able to accurately reproduce our experimental data – including the double-gap structure – across the entire energy range. Since our BTK fits indicate large barrier heights ($Z_1 \geq 0.35$ for $\Delta_1$ and $Z_2 \sim 1000$ for $\Delta_2$), we also attempt to model our data using a two-band Dynes model (with a metallic conduction component to compensate for the lower barrier $Z_1$). Both models yield similar results: at 0.27 K, $\Delta_1 = 6$ meV and $\Delta_2 = 12$ meV from the BTK model, while $\Delta_1 = 4$ meV and $\Delta_2 = 13$ meV from the Dynes fit. $\Delta_1$ exhibits a pronounced anisotropy $\alpha \sim 0.7$, whereas $\Delta_2$ is approximately isotropic ($\alpha = 1$). In each case, $\Delta_2$ provides the dominant contribution to the spectral weight: 60 ± 5 % versus 40 ± 5 % for $\Delta_1$ in the BTK model, compared with a $\Delta_1$:$\Delta_2$:metallic ratio of 8 ± 0.5 % : 46 ± 4 % : 46 ± 4% in the Dynes model. The smaller values for $\Delta_1$ and its spectral weight from the Dynes fit are due to non-negligible Andreev reflections, which cannot be perfectly simulated by the metallic component within this model. We also attempted to reproduce our data using a two-gap *d*-wave model, but no improvement of the fit was observed compared to the anisotropic *s*-wave case.

The large size of $\Delta_2 \geq 12$ meV is a striking feature of our $Bi_2Te_3$/$Fe_{1+y}Te$ heterostructures. It exceeds the superconducting gap in bulk FeSe [24] or $FeSe_{1-x}Te_x$ [21] by at least a factor of 4, despite the $T_c$ of our heterostructures being comparable to $T_c$ in these bulk materials. Our data constitute a demonstration of the potential for strong-coupling superconductivity that could persist up to far higher temperatures than the critical temperatures of the order of 10 K observed in the bulk iron-chalcogenides. The origin of this gap enhancement is unclear, but as the presence of the $Bi_2Te_3$ layer is essential for the appearance of superconductivity here, it is likely that the topological surface states of $Bi_2Te_3$ may play an important role.

Multiband superconductivity with various gaps has been reported in various Fe-based superconductors [31, 38, 39], which is attributed usually to multiple electronic bands crossing the Fermi level in the same material. However, as we are probing the properties of the FeTe and $Bi_2Te_3$ layers in parallel, the twin-gap feature could also originate from two spatially separated regions each with its corresponding electronic bands involved. Superconductivity induced by the proximity effect in the $Bi_2Te_3$ layer could therefore play a role in the opening of the second gap, which appears just below the temperature at which zero resistivity and thus global phase coherence are established. The temperature dependence of the gaps extracted from the BTK and Dynes fits is shown in Figs. 1c,d: it is clear that both models yield qualitatively identical results. Below 2K the smaller gap $\Delta_1$ is approximately constant; upon increasing the temperature its magnitude decreases rapidly, reaching zero between 8 K and 10 K. In contrast, the larger gap $\Delta_2$ is almost temperature-independent between 10 K and 25 K, gradually decreasing towards zero when the temperature is increased further up to 40 K. Between 10 K and 8 K, $\Delta_2$ is slightly reduced, which clearly correlates with the opening of the smaller gap. Below 8 K, the sample exhibits zero resistance and is hence globally phase-coherent. Although the pronounced coherence peaks, which are the characteristic of superconducting tunneling spectra, are absent from our data due to the short quasiparticle lifetimes imposed by low-dimensional fluctuations, we may nevertheless infer the presence of coherence by the sharp falls in *dI/dV* close to $\Delta_{1,2}$. Above 8 K, the global phase coherence is lost and $\Delta_1$ vanishes, while $\Delta_2$ becomes a pseudogap which persists up to 40 K. The Dynes parameter $\Gamma_2$ describing the quasiparticle lifetime in $\Delta_2$ also rises steeply above 8K (Fig. 2e), supporting our observed loss of phase coherence above this temperature. Unfortunately, our data do not allow us to judge whether $\Delta_2$ closes at 8 K and is replaced by a pseudogap due to a competing order, or if it transforms continuously into the high-temperature pseudogap, thus suggesting an at least partially phase-incoherent superconducting (i.e. pairing) origin.

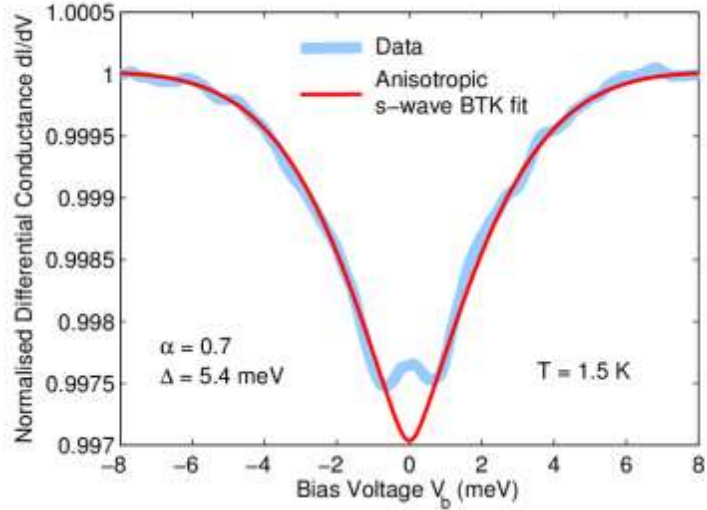

FIG. 2. Zero-field conductance of a point contact on the $Bi_2Te_3$ surface of a $Bi_2Te_3$(9QLs)/$Fe_{1+y}$Te heterostructure, measured by injecting the current perpendicular to the film plane through a tungsten scanning probe tip in contact with the surface. A fit using an anisotropic *s*-wave single-gap BTK model is also shown: $\alpha = 0.7$ and $\Delta = 5.4$ meV, thus corroborating our results for $\Delta_1$ from our edge-contacted heterostructures.

In Fig. 2 we present the differential conductance upon injecting the current through a scanning probe tip perpendicular to the $Bi_2Te_3$ surface. A small 5.4 meV gap dominates the conductance at low temperature, in excellent agreement with the small gap $\Delta_1$ observed upon injecting the current into the edge of the interface. The conductance saturates for $V_b > 6$ mV and no signature of the larger gap $\Delta_2$ is observed. This suggests that the $Bi_2Te_3$ layer becomes superconducting below 8 K and is responsible for $\Delta_1$ in our edge-contacted experiments, similar to what has been observed in $Bi_2Se_3$ films grown on a cuprate high-temperature superconductor [40]. Significantly, a ZBCP is also observed in this spectrum.

  The pseudogap opens well below the antiferromagnetic transition of the bulk $Fe_{1+y}$Te layer and thus does not appear to be directly related to magnetic ordering. Similar to Se-doped FeTe, it is likely that the antiferromagnetism is suppressed by charge transfer across the interface from the *n*-doped $Bi_2Te_3$ layer. Furthermore, the pseudogap does not exhibit any correlation with the upturn of the resistance below 24 K, which is also associated with the bulk layer. Various Fe-based superconductors display a pseudogap well above $T_c$ [18-22, 31] and its origin has been suggested to be nematic quantum critical fluctuations of the antiferromagnetic transition which extend into the normal state [22, 41, 42]. On the other hand, we have previously shown that the interfacial superconductivity in our heterostructures lies in the extreme 2D limit and the resistance drop and *IV* characteristics can be perfectly modelled by a BKT transition [25]. This represents a pure 2D phase-ordering transition of Cooper pairs that are already formed at higher temperature, where the phase of the superconducting order parameter is stabilized below a characteristic temperature $T_{BKT}$ (which occurs in close vicinity above $T_0$) at which thermally-induced vortices and anti-vortices are bound into pairs [43-45]. This naturally implies the existence of phase-incoherent Cooper pairs within a certain temperature range above $T_{BKT}$, creating a pseudogap in the density of states. The 10 meV magnitude of this strongly-coupled superconducting gap indicates a potential for strong coupling superconductivity with stable Cooper pairing at temperatures well above 12 K. Superconducting fluctuations have been observed at temperatures many times higher than $T_c$ in strongly-underdoped layered cuprate HTSCs [10, 12], where they contribute in part to the pseudogap formation. Furthermore, scanning tunneling spectroscopy on ultrathin titanium nitride has shown that the strong phase fluctuations associated with two-dimensionality can induce a pseudogap in conventional superconducting films at temperatures up to 14 times $T_c$ [15]. A superconducting origin for the pseudogap in $Bi_2Te_3$(9QLs)/$Fe_{1+y}Te$ is supported by the fact that the pseudogap is partially suppressed by a magnetic field of 9 T in temperatures up to 40 K (Fig. 3), which could be a consequence of pair-breaking effects. However, in a pseudogap which is entirely caused by fluctuations of the superconducting order parameter, it is expected that the zero-bias conductance $G_N$ should vary linearly with $\ln(\ln(T/T_0))$ [15, 46], where $T_0$ is the phase coherence temperature, represented by the establishment of zero resistivity at 8 K. From Fig. 3b, this trend is not observed in our data, thus suggesting that a competing normal-state pseudogap at least partially contributes to the spectra. A normal state origin is furthermore compatible with the increase in the larger gap $\Delta_2$ above 8 K, which is suggestive of a replacement of the superconducting gap by a normal state pseudogap.

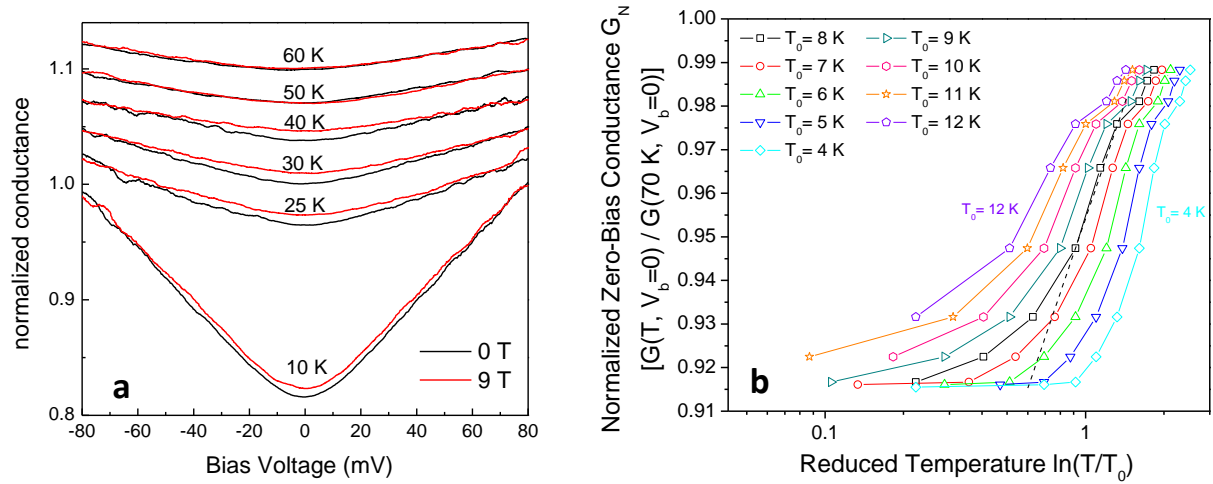

FIG. 3a Pseudogap in the normalized spectra in zero field and 9 T, illustrating a partial suppression of the pseudogap by the applied field at temperatures below 40 K. **b** Temperature dependence of the normalized zero-bias conductance $G_N$ above the phase-coherence temperature $T_0$ as a function of the reduced temperature $\varepsilon = \ln(T/T_0)$ for various choices of $T_0$ surrounding our experimentally-determined $T_0 = 8$ K (the dotted line is a guide to the eye). The absence of linear behavior over a sufficiently large range of $\varepsilon$ suggests a competitive normal state origin for the pseudogap rather than phase-incoherent pairing [15, 46].

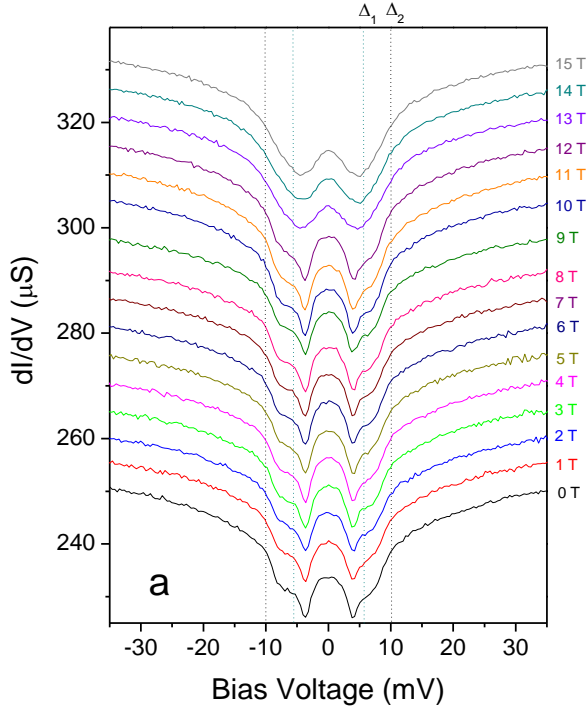

FIG. 4. Point-contact differential conductance on the edge of a $Bi_2Te_3$(9QLs)/$Fe_{1+y}Te$ heterostructure at 0.27 K in various magnetic fields applied perpendicular to the interface. An offset has been added to each spectrum for clarity; dotted lines indicate constant energies and are guides to the eye.

In Fig. 4, we plot the point-contact spectra in magnetic fields up to 15 T applied perpendicular to the film plane. The gap structure and ZBCP are quite robust with respect to magnetic fields: at first glance, the spectra hardly change up to 12 T. In higher fields the gap becomes shallower, but the gap energies $\Delta_{1,2}$ do not shrink significantly. This resilience of the overall gap structure demonstrates that the Cooper pairing strength is almost impervious to strong fields, in direct contrast with the critical field $H_{c2} = 17$ T which has been estimated from resistance data [25]. We deduce that $H_{c2}$ merely corresponds to a field-induced loss of phase coherence.

As demonstrated by our point contact on the top of the $Bi_2Te_3$ layer, interfacial contact with the $Fe_{1+y}Te$ induces superconductivity in $Bi_2Te_3$, which is therefore likely to host a topological superconducting state [50]. The combination of superconductivity with the non-trivial topological symmetry of the surface states in a TI naturally evokes the question whether the ZBCP could be caused by Majorana bound states [47, 48, 49, 50]. Care has been taken to eliminate any spurious origin for the ZBCP, e.g. heating or proximity effects [51]: (1) our high-resistance point-contact lies comfortably within the ballistic spectroscopic tunneling regime, (2) the spectra were verified to be identical upon increasing and decreasing the tunnel current and (3) the ZBCP width remains roughly constant at all temperatures below 8 K (Fig. 1a), despite the gap energy $\Delta_1$ increasing from zero to 5 meV within this temperature range. The ZBCP is therefore of intrinsic origin. The proximity effect of an s-wave superconductor on a TI [27] would cause a fully gapped energy spectrum without any in-gap states. Majorana modes can then only be created in the vortex cores in an applied magnetic field. Since the pairing symmetry in $Bi_2Te_3$(9QLs)/$Fe_{1+y}$Te remains unclear, two possible mechanisms exist for the formation of Majorana edge states. One possibility is that the contact with FeTe drives the $Bi_2Te_3$ layer to become intrinsically a topological superconductor with Majorana surface states [47, 50, 52-54], .e.g. by a charge transfer effect, instead of a proximity-induced superconductivity. This would explain the ZBCP observed in the point contact spectra of both the edge (Fig. 1) and the top contacts (Fig. 2). In a magnetic field the ZBCP should be suppressed due to broken time-reversal symmetry: a reduction in the ZBCP height is indeed observed in high magnetic fields, but the effect is only ~20% in 15 T applied in-plane and perpendicular to the interface. This could be a consequence of a particularly strong Rashba spin-orbit coupling at the interface as well as of the extremely high critical field for pairing. A detailed theoretical study of the possible pairing phases of $Bi_2Te_3$ would be required to confirm this possibility. The alternative mechanism requires nodal $d_{x^2-y^2}$ superconductivity (associated with the $Fe_{1+y}$Te) combined with Rashba spin-orbit coupling (enhanced by the topological surface states of $Bi_2Te_3$) [55]. Although a *d*-wave order parameter alone could create a fermionic ZBCP at the sample edge, a ZBCP composed of the fermionic edge states and Majorana fermions should form in the presence of strong spin-orbit coupling at zero field. It has been predicted that an in-plane applied field will split and shift the fermionic states to finite energy via the Zeeman effect, while the Majorana state remains at zero energy [55]. However, no splitting of the ZBCP is observed up to 15 T for applied parallel or perpendicular to the interface, and its height is rather small, thus rendering such a *d*-wave scenario unlikely. Furthermore, the ZBCP visible in our top contact (Fig. 2) is at odds with a *d*-wave scenario.

A more conventional explanation for the ZBCP is related to the large Fe excess in the $Fe_{1+y}$Te layer: scanning probe measurements have recently shown that excess iron in Fe(Te,Se) superconductors causes pronounced local in-gap states, which do not split, shift or vanish in applied magnetic fields [56]. However, although Fe impurities could certainly explain the presence of in-gap states in our heterostructures, it would be rather surprising if they appeared at exactly zero bias. The presence of a ZBCP when tunneling directly into the $Bi_2Te_3$ layer (in which Fe impurities are absent) also discourages this interpretation. The origin of the ZBCP therefore remains a mystery and requires further experiments such as angle resolved photoemission to clarify the exact electronic density of states in $Bi_2Te_3$/$Fe_{1+y}$Te heterostructures. Our $Bi_2Te_3$(9QLs)/$Fe_{1+y}$Te heterostructures reveal a highly unusual superconducting state with an extraordinarily large superconducting gap, a pronounced pseudogap and compelling evidence for proximity-induced superconductivity in the topological insulating $Bi_2Te_3$ top layer. Our techniques cannot yet determine the origin of these features indisputably; however, it is clear that our data display a similarly rich behavior to the cuprate superconductors, which remain one of the major unsolved mysteries in physics.

We thank M. L. Cohen, Y.-R. Shen and S. G. Louie for stimulating discussions and U. Lampe for technical support. This work was supported by grants from the Research Grants Council of